\documentclass[11pt,bibliography=totoc]{scrartcl}

\usepackage{soul}
\usepackage[utf8]{inputenc}
\usepackage[english]{babel}
\usepackage{graphicx,amsmath,amsfonts,amssymb,dcolumn,amsthm,slashed,bbm,xspace,pgffor,tikz,mathbbol,latexsym,cite,braket}
\usepackage{microtype}
\usepackage{mathtools}
\usepackage{nccmath}
\usepackage{lmodern}
\usepackage[normalem]{ulem}
\usepackage{xcolor}
\usepackage[colorlinks=true,citecolor=blue,urlcolor=blue,linkcolor=blue]{hyperref}
\usepackage[hang]{footmisc} 
\usepackage[affil-it]{authblk}
\usepackage{multicol}

\numberwithin{equation}{section}

\begin{document}

\title{\textbf{Configuration entropy of a rotating quark-gluon plasma from holography 
}}

\author{Nelson R.~F.~Braga$^{a}$\thanks{\href{mailto:braga@if.ufrj.br}{ braga@if.ufrj.br}} , ~ Luiz F.~Ferreira$^{b}$\thanks{\href{mailto:lffaulhaber@mgail.com}{ lffaulhaber@mgail.com}}  ,~  Octavio C.~Junqueira$^a$\thanks{\href{mailto:octavioj@pos.if.ufrj.br}{octavioj@pos.if.ufrj.br}} }
\affil{ $^{a}$ UFRJ --- Universidade Federal do Rio de Janeiro, Instituto de Física,\\
Caixa Postal 68528, Rio de Janeiro, Brasil}

\affil{ $^{b}$ Instituto de Física y Astronomia 
,\\ 
Universidad de Valparaiso, \\
A. Gran Bretana 1111, Valparaiso, Chile}

\date{}
\maketitle

\begin{abstract}

The configuration entropy (CE) provides a measure of the stability of physical systems that are spatially localized. An increase in the CE is associated with an increase in the instability of the system. In this work we apply a recently developed holographic description of a rotating plasma,  in order to investigate the behaviour of the CE when the plasma has angular momentum.  Considering the  holographic dual to the plasma, namely a rotating AdS black hole, the CE is computed at different rotational speeds and temperatures. The result obtained shows not only an increase with the rotational speed $ v$ but, in particular, a divergence of the CE as $v$  approaches the speed of light:  $\, v   \to 1 $. We discuss an interpretation for the increase in the CE in terms of the emission of radiation from the black hole and from its dual plasma.
 
\end{abstract}

\section{Introduction}

In the last decades, one of the focus of the community devoted to studying the physics of  strong interactions is the search for understanding the Quark-Gluon Plasma (QGP), a state of matter formed by deconfined partons that  interact strongly\cite{Busza:2018rrf}. 
The QGP  is formed experimentally through ultra-relativistic collisions of heavy  nuclei produced in  particle accelerators. There are many properties that affect the behaviour of QGP, like temperature, density and the presence of magnetic fields. Another, less studied, property is the rotation of the plasma. When the collision is non-central, the resulting system  acquires not only a strong magnetic field but also angular momentum. A fraction of this angular momentum  can be transferred to the  polarization of the strange quarks  in the QGP due to spin-orbit interaction, as discussed in \cite{Liang:2004ph}. Until recently, no experimental signals of this effect had been detected.  Only in 2017, the  global hyperon polarization has been observed in $Au+Au$ collisions at RHIC \cite{STAR:2017ckg}. This discovery  opened a new window to study the properties of rotating Quantum Chromodynamics (QCD) matter. In particular, the influence of the rotation on the QCD phase diagram have become a topic under active investigations. Some recent  studies of the effect of rotation in the QGP can be found, for example, in Refs.  \cite{Miranda:2014vaa,Jiang:2016wvv,McInnes:2016dwk,Mamani:2018qzl,Wang:2018sur,Chernodub:2020qah,Arefeva:2020jvo,Chen:2020ath,Zhou:2021sdy,Braguta:2021jgn,Braguta:2021ucr,Golubtsova:2021agl,Fujimoto:2021xix,Braga:2022yfe,Chen:2022mhf,Golubtsova:2022ldm,Zhao:2022uxc,Chernodub:2022veq}. 

In particular, using the AdS/QCD approach  \cite{Chen:2020ath,Braga:2022yfe}, it was found that the critical temperature of the confinement/deconfinement transition decreases with increasing angular velocity, which is in agreement with others phenomenological models as the Nambu-Jona Lasinio (NJL). In contrast,  the simulations of relativistic rotation on the confinement/deconfinement phase transition in gluodynamics lattice \cite{Braguta:2021jgn,Braguta:2021ucr} showed that the critical temperature increases  with increasing angular velocity.

Here we are interested in studying how rotation affects the stability of the quark-gluon plasma. So, it is important to make it clear what does one mean by stability in this case. 
There are two aspects to be considered. The first is the fact that the plasma phase of QCD matter exists, or is stable, for temperatures above some critical value $T_c$. For lower temperatures,  QCD matter is in the confined phase\cite{Herzog:2006ra,BallonBayona:2007vp}. The other instability is related to the Hawking radiation. The higher is the temperature of the black hole dual to the plasma, the stronger is the emission of radiation and the corresponding loss of energy, resulting in an increase in the instability. 

A rotating plasma with uniform rotational speed is described holographically by a rotating black hole with cylindrical symmetry, like those studied in \cite{BravoGaete:2017dso, PhysRevD.97.024034}. The rotation is obtained by a coordinate transformation and the holographic model obtained predicts that plasma rotation affects the critical temperature of confinement/deconfinement transition \cite{Braga:2022yfe}. This fact motivated the present study concerning the stability of the rotating plasma, but now considering the configuration entropy (CE) approach.

The idea of using the CE as an indicator of stability of physical systems started in \cite{Gleiser:2011di,Gleiser:2012tu,Gleiser:2013mga}. Afterwards many examples appeared in the literature where the CE plays the role of representing stability in different  physical systems. For instance, in the context of AdS/QCD approach, the CE has provided new results  involving   thermal behaviour of quarkonium in a medium \cite{Braga:2018fyc,Braga_2018,Braga:2020myi,Braga:2020hhs,Braga:2021fey}, the mass spectra of several particles  \cite{Brenardini_2017,Bernardini:2018uuy,MarinhoRodrigues:2020ssq,daRocha:2021ntm,Ferreira:2019nkz,Ferreira:2019inu,Ferreira:2020iry}, nuclear electromagnetic transitions \cite{daRocha:2022bnk} and in confinement/deconfinement transition \cite{Braga:2020opg,Lee:2021rag,Braga:2021zyi}. In particular, in ref. \cite{Braga:2021fey} it was shown how does the CE marks the total dissociation of quasi-particles in a medium. Namely, the CE diverges when the particles are completely dissociated.  Other applications in astrophysics, cosmology, AdS/CFT correspondence, nuclear physics and field theory  can be found, for example,  in references 
\cite{Correa:2015vka,Braga:2016wzx,Karapetyan:2016fai,Karapetyan:2017edu,Karapetyan:2018oye,Lee:2018zmp,Bazeia:2018uyg, Braga:2019jqg,Stephens:2019tav,Alves:2020cmr,daRocha:2021jzn,Casadio:2022pla,Ma:2022lox,Barreto:2022ohl}. 

The CE is based on the Shannon information entropy \cite{shannon}, in the continuum limit. It is in general defined in terms of the energy density of the physical state in Fourier space. Assuming that the black hole is represented by the grand canonical ensemble, which is consistent with the fact that the plasma is rotating, one can find an expression for the rotating black hole energy density as a function of the holographic coordinate, the temperature and the rotational speed. The Fourier transform does not possess an analytical solution, so that the results in this work were obtained through the application of numerical methods.  

The results obtained show a monotonic increase of the CE with the
rotational speed for a fixed temperature. In particular, the CE diverges as the rotational speed of the black hole approaches the asymptotic limit corresponding to the speed of light. 

This work is organized as follows: in Section 2, we describe the geometry of a rotating cylindrical black hole. Section 3 is devoted to the construction of the rotating plasma energy density, in the soft wall AdS/QCD model, assuming that the system is represented by the grand canonical ensemble. In Section 4, we compute the CE of the rotating plasma at different rotational velocities and temperatures, and discuss an interpretation for the results obtained.  Section 5 contains our conclusions and the appendix shows a table with additional results for the CE as a function of the temperature and rotational speed.


\section{Rotating AdS black hole with cylindrical symmetry}

\subsection{AdS black hole}
At finite temperature, asymptotically anti-de Sitter spaces with negative constant curvature possesses two solutions given by the following metrics with compact time direction:
\begin{equation}\label{thermalAdS}
    ds^2= \frac{L^2}{z^2}\left( - dt^2 + d\overrightarrow{x}^2 + dz^2 \right)\;,
\end{equation} 
and 
\begin{equation}\label{BHAdS}
ds^2= \frac{L^2}{z^2}\left( - f(z) dt^2 + d\overrightarrow{x}^2 + \frac{dz^2}{f(z)} \right)\;,
\end{equation}
with $f(z) = 1 - z^4/z_h^4$. The first geometry \eqref{thermalAdS} corresponds to the thermal AdS space, while the second one \eqref{BHAdS} to the AdS black hole (BH) geometry, being $z_h$ the location of the horizon. Both geometries are solutions of Einstein's equations with negative cosmological constant $\Lambda = - 12/L^2$ and constant curvature $R = -20/L^2$. 

The time coordinate of the BH geometry has a period $\beta$, related to the horizon position and to the Hawking temperature,  $T = 1/\beta = 1/(\pi z_h)$ \cite{Hawking:1982dh}. Requiring that the asymptotic  limits of the two geometries at $z=\epsilon$, with $ \epsilon \to 0 $,  are  the same, one finds that the period of the time component of the thermal AdS space is $\beta^\prime = \pi z_h \sqrt{f(\epsilon)}$. 

\subsection{ Cylindrical AdS black hole}

In this work we will consider, as a first step in understanding  the effect of rotation in the QGP, the simple case when all the points have the same rotational speed. With this purpose, we make a change in the metrics \eqref{thermalAdS} and \eqref{BHAdS}, corresponding to the compactification of coordinate $x_3$, that leads to the cylindrical geometries, respectively
\begin{equation}\label{cthermalAdS}
	ds^2 = \frac{L^2}{z^2} \left(- d{\bar t}^2 + l^2 d\bar \phi^2 + \sum_{i=1}^2 dx_i^2 + dz^2\right)\;, 
\end{equation}
and
\begin{equation}\label{CBH}
	ds^2 = \frac{L^2}{z^2} \left(- f(z)d{\bar t}^2 + l^2 d\bar \phi^2 + \sum_{i=1}^2 dx_i^2 + \frac{dz^2}{f(z)}\right)\;, 
\end{equation}
where  $l$ is  the radius of a hyper-cylinder and $ 0 \le \bar \phi \le 2 \pi $. Note that  these new metrics have a topology that is different from those of metrics \eqref{thermalAdS} and \eqref{BHAdS} since one of the spatial coordinates has been compactified. However, the curvature of spacetime is the same as that of eqs. \eqref{thermalAdS} and \eqref{BHAdS}. So it makes sense to call these metrics as cylindrical BH AdS and thermal AdS. Then, it is natural to take \eqref{CBH} as the gravity dual of a (hyper)cylindrical QGP at rest. Or, in other words, as seen by an observer that is at rest with respect to the QGP. 

The change to cylindrical symmetry does not modify the relation between the BH temperature and the horizon position:
\begin{equation}
T =   \frac{1}{\pi z_h } \, .
\end{equation}

\subsection{Cylindrical black hole as observed by a rotating frame}
Now we perform a coordinate transformation to an observer $O$ relative to which the observer $ \bar O$ of the previous subsection is rotating uniformly. In other words, we perform a coordinate change such that for a fixed value of the original angular coordinate $\bar \phi $ the new one $\phi $ varies uniformly:

\begin{eqnarray}
	{\bar t} & = & \frac{1}{\sqrt{1-l^2 \omega^2}} \left(t + l^2 \omega \phi \right)\;,\label{T1}\\
	{\bar \phi} & = & \frac{1}{\sqrt{1-l^2 \omega^2}} \left(\phi +  \omega t \right)\label{T2}\;,
\end{eqnarray}
These coordinate transformations lead to the new metric
for the BH \cite{Zhou:2021sdy}
\begin{eqnarray}\label{RotatingBHmetric}
	ds^2 &= & g_{tt} dt^2 + g_{t\phi}dt d\phi + g_{\phi t} d\phi dt + g_{\phi \phi} d\phi^2 \cr && + \,  g_{zz} dz^2 + g_{xx} \sum_{i=1}^2 dx_i^2\;,
\end{eqnarray}
with
\begin{eqnarray}
	g_{tt} &=& \frac{\gamma^2(\omega)L^2}{z^2}\left( \omega^2 l^2-f(z) \right)\;, \label{RotatingBHi} \\
	g_{\phi \phi} &=& \frac{\gamma^2(\omega)L^2}{z^2} l^2 \left( 1-\omega^2 l^2f(z) \right)\;,\\
	g_{t \phi} &=& g_{\phi t} = \frac{\gamma^2(\omega)L^2}{z^2}\omega l^2 \left( 1-f(z) \right)\;,\\
	g_{zz} &=& \frac{L^2}{z^2 f(z)}\;,\\
	g_{xx} &=& \frac{L^2}{z^2}\;,\label{RotatingBHf}
\end{eqnarray}
where $\gamma$ is the Lorentz factor,
\begin{equation}
	\gamma(\omega l ) = \frac{1}{\sqrt{1-l^2\omega^2}}\;.
\end{equation}

Notice that the new angular variable $\phi $ has a different periodicity interval: $ 0 \le   \phi \le 2 \pi /\gamma $. In order to compute the contribution of this factor to the volume of space-time, we must change again to an angular coordinate $ \phi^\prime = \phi \gamma $. This coordinate has the same periodicity   $ 0 \le   \phi^\prime  \le 2 \pi   $ of the original one, $ \bar \phi $. Then, if one calculates the corresponding new metric components and metric determinant $g^\prime$, one finds $g^\prime = \frac{l^2 L^{10}}{z^{10}\gamma^2} = g/\gamma^2$, meaning  that the space-time volume is changed by a factor $1/\gamma$ after the transformations \eqref{T1} and \eqref{T2}. (This result will be used  when we integrate in $\phi $ to calculate the action integrals in the next section.)  

 The geometry corresponding to the metric \eqref{RotatingBHmetric} represents, through holography, a plasma of cylindrical form as observed by a frame that is in uniform rotation with respect to it. An interesting way to clarify the fact that the coordinate transformations \eqref{T1} and \eqref{T2} lead to the description of a rotating plasma is to analyse the angular momentum $J$. For the metric \eqref{RotatingBHmetric} $J$ can be calculated using holographic renormalization \cite{Braga:2022yfe} (see also \cite{PhysRevD.97.024034,BravoGaete:2017dso}) with the result:
 \begin{equation}
J = -\frac{\partial \Phi_G}{\partial \omega}=\frac{2L^3 }{\kappa^2}\frac{\omega}{z_{h}^{4}(1-w^2l^2)}\;,
\end{equation} 
For the original metric \eqref{CBH}, before the coordinate transformations, one has $ J = 0$. So, the
effect of the coordinate transformations is indeed that of adding angular momentum to the system and consequently representing a plasma in rotation. 
 
  As shown in \cite{Braga:2022yfe}, the metric \eqref{canon} is a solution of the same Einstein equation satisfied by the AdS black hole metrics \eqref{BHAdS} and \eqref{CBH}. For the cylindrical thermal AdS space of eq. \eqref{cthermalAdS}, the rotating version of the metric has the same form of \eqref{canon} but using $f (z)  = 1 $ in eqs. \eqref{N}-\eqref{P}.

 Its is important to stress the fact that we are considering here the very simple case of a cylindrical shell, assumed to be in equilibrium, observed in rotation. All the points move with the same speed. In the realistic case of the QGP formed in heavy ion collisions, the rotational speed is clearly not uniform. So, in the real QGP there are some important issues that we are not considering here related, for example with the interaction among layers of the plasma that have different velocities. The condition for equilibrium in the case of a rotating plasma is non-trivial. 

 Our aim is to gain some understanding of the qualitative effects of the plasma rotation. This approach has some similarity with the one used for studying the effect of magnetic fields on the plasma, where in general it is considered that the field is uniform \cite{Dudal:2014jfa,BRAGA2018186,BRAGA2019462,BRAGA2020135918,Braga:2021fey, PhysRevD.103.086021, Bohra:2019ebj, Bali:2011qj, PhysRevD.86.016008, Ballon-Bayona:2013cta, Li:2016gfn, Ballon-Bayona:2017dvv}, although the actual  fields acting on the QGP formed in  non-central heavy ion collisions are not uniform. 


It is important to note that there is a condition to be satisfied by the velocity. Namely: $\omega l < 1 $. On the other hand, in this work the relevant quantity is $\omega l  $ and not the radius or the angular velocity separately. However, if one wants to superimpose different regions of a plasma that rotates with the same angular speed but with different radius than these quantities would have independent rules.

After a straightforward calculation, one can rewrite the rotating cylindrical black hole metric \eqref{RotatingBHf}-\eqref{RotatingBHf} in the canonical form:
\begin{eqnarray}\label{canon}
	ds^2 &=& -N(z) dt^2 + \frac{L^2}{z^2}\frac{dz^2}{f(z)} + R(z)\left( d\phi^\prime + P(z) dt\right)^2    + \frac{L^2}{z^2} \sum_{i-1}^2 dx_i^2\;,
\end{eqnarray}
with 
\begin{eqnarray}
	N(z) &=& \frac{L^2}{z^2} \frac{ f(z) (1-\omega^2 l^2)}{1- f(z)\omega^2 l^2}\;, \label{N}\\
	R(z) &=& \frac{L^2}{\gamma z^2}\left( \gamma^2 l^2 -  f(z) \gamma^2 \omega^2 l^4\right)\;, \label{R}\\
	P(z) &=& \frac{\omega\gamma(1-f(z))}{1- f(z)\omega^2 l^2}\;\label{P}.
\end{eqnarray}
Thus, defining $h_{00} = - N(z)$, the  Hawking temperature can be obtained from the surface gravity formula \cite{Zhou:2021sdy}:

\begin{eqnarray}\label{HT}
	T &=& \vert \frac{\kappa_G}{2\pi} \vert = \bigg\vert \frac{\lim_{z\rightarrow z_h}- \frac{1}{2} \sqrt{\frac{g^{zz}}{-h_{00}(z)}}h_{00,z}}{2\pi}\bigg \vert = \frac{1}{\pi z_h} \sqrt{1-\omega^2 l^2}\;, 
\end{eqnarray}
where $\kappa_G$ is the surface gravity, and $g^{zz}$ the $zz$ component of the inverse of the cylindrical black hole metric. The expression of Eq. \eqref{HT} shows that the rotation affects the relation between the Hawking temperature $T$ of the black hole and its horizon position. 
It was shown in \cite{Braga:2022yfe} that the condition for the occurrence of the Hawking Page (HP)  transition depends only on the horizon position $z_h$, not on the temperature.  Equation \eqref{HT} tells us that, for a fixed temperature, an increase in the rotational speed corresponds to a decrease in $z_h$ by a factor of $ \sqrt{1-\omega^2 l^2} $. Or, equivalently, the fixed value of $z_h$ where the 
HP transition occurs will correspond to a temperature that decreases with $ \frac{1}{ \sqrt{1-\omega^2 l^2} }$ as $wl $ increases. Therefore, regarding the HP transition the plasma becomes more stable when the rotational speed increases. 

However, as we mentioned in the introduction, a black hole can be unstable also because of the Hawking radiation. We will discuss this fact when interpreting the results for the CE in section \textbf{4}.

\section{Energy density in the soft wall AdS/QCD model}

The soft wall holographic AdS/QCD model  \cite{Karch:2006pv}  is built from the introduction in the AdS geometry of a dilaton background $ \Phi (z)=cz^2  $, where $c$  is a parameter with dimension of energy squared.   In the gauge theory side of the gauge/gravity duality $\sqrt{ c} \, $ plays the role of an infrared energy  (IR) cutoff.

\subsection{Regularized action density}

One can write the five-dimensional gravitational action, at zero temperature,  in the general form\cite{Herzog:2006ra,BallonBayona:2007vp}
\begin{equation}\label{action1}
	I = - \frac{1}{ 2 \kappa^2} \int_0^{z_f} dz\int d^4x \sqrt{- g} e^{-\Phi}\left( R - \Lambda \right) \; =  \frac{4}{L^2\kappa^2} \int_0^{z_f} dz \int d^4x \sqrt{- g} e^{-cz^2},     
\end{equation}
where $\kappa$ is the gravitational coupling associated with the Newton constant, $ \sqrt{c} $ is the IR energy parameter and $\Lambda$ the cosmological constant, related to the curvature as  $\Lambda = \frac{3}{5} R = \frac{-12}{L^2}$. The integral in $z$ has an upper limit, that we named as $z_f$, that is equal to $z_h$ for the black hole case and to $ z \to \infty $ for the thermal AdS case. 

In order to compute the total energy density in the grand canonical ensemble and perform the analysis of the Hawking-Page transition, we must determine the on-shell regularized action. From the rotating metric \eqref{RotatingBHmetric} and the transformation $\phi^\prime = \gamma \phi$, the determinant of $g^\prime_{\mu\nu}$ is  $g^\prime = \frac{l^2L^{10}}{z^{10} \gamma^2}$ (see previous section), so that  Eq. \eqref{action1}  becomes
\begin{equation}
	I_{\text{on-shell}} = \frac{4L^3}{\kappa^2\gamma} \int d^3x \int_0^\beta dt\int_0^{z_f}  dz \, z^{-5}e^{-cz^2}\;.
\end{equation} 
Since the integration over the spatial bulk coordinates $d^3x$ is trivial and only contributes with a volume factor, $V_{3D}$, we now define an action density $\mathcal{E} = \frac{1}{V_{3D}} I_{\text{on-shell}}$. For a compact time direction in the Euclidean signature, we have $ 0 \leq t < \beta_s $, where $\beta_s $ depends on the space considered and the general form of the action density is then given by the expression
\begin{equation}
	\mathcal{E}_s(\varepsilon) = \frac{4L^3}{\kappa^2\gamma} \int_0^{\beta_s}dt \int_\varepsilon^{z_f}dz\, z^{-5}e^{-cz^2}  \;,  
\end{equation}
where $\varepsilon$ is an ultraviolet (UV) regulator. As mentioned before, for the black hole geometry, $ \beta_{BH} = 1/T$, while for the thermal AdS one, $ \beta_{AdS}  = \sqrt{f(\epsilon)} \beta_{BH} = \sqrt{f(\epsilon)} / T $. 

We must introduce the UV regulator because the black hole and thermal AdS space possesses infinite action densities. In order to eliminate these divergencies, one defines the regularized action density of the rotating black hole as the difference between the action densities of the two geometries,
\begin{equation}\label{DeltaE}
	\bigtriangleup \mathcal{E}(\varepsilon) = \lim_{\varepsilon \rightarrow 0} \left[\mathcal{E}_{BH}(\varepsilon) - \mathcal{E}_{AdS}(\varepsilon) \right]\;, 
\end{equation}
where
\begin{eqnarray}
	\mathcal{E}_{BH}(\varepsilon) &=& \frac{4L^3}{\kappa^2\gamma} \beta \int_\varepsilon^{z_h}dz\, z^{-5}e^{-cz^2}  \;, \label{DeltaEBH}\\
	\mathcal{E}_{AdS}(\varepsilon) &=& \frac{4L^3}{\kappa^2\gamma} \sqrt{f(\epsilon)} \, \beta \int_\varepsilon^{\infty}dz\, z^{-5}e^{-cz^2}  \;. \label{DeltaEAdS}
\end{eqnarray}

From the regularized action density \eqref{DeltaE}, one can determine the energy density of the rotating BH in the grand canonical ensemble, taking into account the contribution of the angular momentum. From gauge/gravity duality, it will correspond to the energy density of the rotating plasma with cylindrical symmetry.

\subsection{Total energy density in the grand canonical ensemble}

The first step to compute the configuration entropy  of the plasma at different temperatures and angular velocities is to determine the total energy density. 
First we consider the total energy $E$. Due to rotation, we should assume that the system is represented by the grand canonical ensemble. In this case, the total energy, with zero chemical potential, is defined by 
\begin{equation}
E = - \frac{ \partial \log Z}{\partial \beta} + \omega J\;, 
\end{equation}
where $Z$ is the partition function, and J, the angular momentum, which in turn is defined by
\begin{equation}
J = \frac{1}{\beta} \frac{\partial \log Z}{\partial \omega}\;. 
\end{equation}
In the semiclassical Hawking-Page approach, $\log Z = -I$, such that 
\begin{equation}\label{totalE}
E =  \frac{ \partial I }{\partial \beta} + \omega J\;, \quad \text{with} \quad J = -\frac{1}{\beta} \frac{\partial  I}{\partial \omega}\;.
\end{equation}
As it was done in the previous subsection, we factorize the trivial three dimensional volume in the   bulk spatial variables $V_{3D}$. This corresponds to replacing in the previous Eqs. the action $I$ by $\Delta \epsilon $. Explicitly: 
\begin{equation}\label{totalE/DV}
U \,\equiv \, \frac{E}{V_{3D}} =  \frac{ \partial \bigtriangleup \mathcal{E} }{\partial \beta} + \omega \,  \frac{1}{\beta} \frac{\partial \bigtriangleup \mathcal{E} }{\partial \omega}\;.
\end{equation}
Now we define an energy density in the holographic coordinate $z\,$,  
$\rho_{BH}(z, T, \omega)$, by:
\begin{equation}\label{rhoBHint}
\int_0^\infty \rho_{BH}(z, T, \omega) \, dz =  U(T, \omega) \;. 
\end{equation}

Thus, using equations \eqref{DeltaE}, \eqref{DeltaEBH} and \eqref{DeltaEAdS}, together with
\begin{equation}
\beta \sqrt{f(\epsilon)}  =  \beta - \frac{\pi^4\epsilon^4}{2\beta^3 (1-\omega^2l^2)^2}\;,
\end{equation}
and introducing the coordinate $u = z/z_h$, one can write \begin{eqnarray}\label{I(u)}
\bigtriangleup \mathcal{E}  &=&  \frac{4L^3}{\kappa^2\gamma} \left[  \beta \int_{\epsilon \pi/(\beta\sqrt{1-\omega^2l^2})}^{1} du\, u^{-5}\left(\frac{\beta\sqrt{1-\omega^2 l^2}}{\pi}\right)^{-4} e^{-c(\frac{u\beta\sqrt{1-\omega^2 l^2}}{\pi})^2}\right.\nonumber\\
&-& \left. \left( \beta - \frac{\pi^4\epsilon^4}{2\beta^3(1-\omega^2 l^2)^2}\right) \int_{\epsilon \pi/(\beta\sqrt{1-\omega^2 l^2})}^\infty du\, u^{-5} \left(\frac{\beta\sqrt{1-\omega^2 l^2}}{\pi}\right)^{-4} e^{-c(\frac{u\beta\sqrt{1-\omega^2l^2}}{\pi})^2}   \right]\;.
\end{eqnarray}
Taking the derivative of Eq. \eqref{I(u)} with respect to $\beta$, and eliminating terms $\mathcal{O(\epsilon)}$ by taking the ultraviolet limit $\epsilon \rightarrow 0$, the first term of the total energy, $U_1 = \partial \bigtriangleup \mathcal{E} /\partial \beta$  in  Eq. \eqref{totalE/DV}, reads 
\begin{eqnarray}\label{RegV7}
U_1 &=&  \frac{4L^3}{\kappa^2\gamma} \left[ - \frac{7\epsilon^4}{2z_h^8} \int_{\epsilon \pi/(\beta\sqrt{1-\omega^2 l^2})}^{1} du\, u^{-5}e^{-c(\frac{u\beta\sqrt{1-\omega^2 l^2}}{\pi})^2} +\frac{1}{2 z_h^4} \right. \nonumber\\
&+&\left.\frac{3}{z_h^4}  \int_1^\infty du\,u^{-5} e^{-c(\frac{u\beta\sqrt{1-\omega^2 l^2}}{\pi})^2}
+  \frac{2c}{z_h^2} \int_1^\infty du\, u^{-3}e^{-c(\frac{u\beta\sqrt{1-\omega^2 l^2 }}{\pi})^2} \right]\;.
\end{eqnarray}
The first two terms of the equation above can be rewritten in the form 
\begin{equation} 
- \frac{7\epsilon^4}{2z_h^8} \int_{\epsilon \pi/(\beta\sqrt{1-\omega^2 l^2})}^{1} du\, u^{-5}e^{-c(\frac{u\beta\sqrt{1-\omega^2 l^2}}{\pi})^2} + \frac{1}{2z_h^4} = -\frac{3}{8z_h^4}\;,
\end{equation}
then, using the following Gaussian integration, 
\begin{equation}\label{G1}
\lim_{\epsilon \rightarrow 0} \left[ -\frac{3\sqrt{c}}{4 z_h^3 \sqrt{\pi}} \int_{\epsilon z_h}^\infty e^{-c (u z_h)^2}\, du\right] = -\frac{3}{8z_h^4} \;, 
\end{equation}
and returning to the holographic $z$-variable, one obtains
\begin{equation}
U_1  = \lim_{\epsilon \to 0}  \left(  \int_{\epsilon}^{z_h} \rho^{(1)}_1(z) \,dz + \int_{ z_h}^{ \infty }\rho_1^{(2)}(z) \,dz  \right) \;,
\end{equation} 
with
\begin{eqnarray}\label{density1}
\rho^{(1)}_1(z) &=& - \frac{4L^3}{\kappa^2\gamma} \frac{3\sqrt{c}}{4 z_h^4 \sqrt{\pi}} e^{-cz^2}  \;, \quad \,\,\,\,\;\;\,\quad\quad\quad\quad \quad\, \text{if} \quad \epsilon  \leq z \leq z_h\;, \nonumber\\
\rho^{(2)}_1(z) &=& \frac{4L^3}{\kappa^2\gamma} \left(-\frac{3\sqrt{c}}{4 z_h^4 \sqrt{\pi}}+ \frac{3 }{z^5}  +  \frac{2 c }{z^3} \right) e^{-cz^2} \;, \quad \text{if} \quad z \geq z_h\;.
\end{eqnarray}

In order to compute the second term of the total energy, 
$U_2 =  \frac{1}{\beta} \frac{\partial \bigtriangleup \mathcal{E} }{\partial \omega}$, we must  work out the expression 
 
\begin{eqnarray}\label{RegV8}
\frac{\partial \bigtriangleup \mathcal{E} }{\partial \omega} &=&  -\frac{4L^3}{\kappa^2 \gamma} \left[  \frac{4\epsilon^4l^2\omega}{z_h^8(1-l^2\omega^2)} \int_{\epsilon/z_h}^{1} du\, u^{-5}e^{-c(uz_h)^2} - \frac{l^2\omega}{2 z_h^4(1-l^2\omega^2)}\right. \nonumber\\
&-&\left. \frac{4l^2\omega}{z_h^4(1-l^2\omega^2)} \int_1^{\infty} du\, u^{-5}e^{-c(uz_h)^2} -  \frac{2c l^2\omega}{z_h^2(1-l^2\omega^2)} \int_1^{\infty} du\, u^{-3}e^{-c(uz_h)^2}\right]\;.
\end{eqnarray}
Similarly to the previous calculation to determine $U_1$, by using the expression
\begin{equation}\label{G2}
\frac{l^2 \omega}{2 z_h^4(1-l^2\omega^2)} = \frac{l^2\omega \sqrt{c}}{z_h^3\sqrt{\pi}(1-l^2\omega^2)} \int_{\epsilon/z_h}^\infty e^{-c(u z_h)^2}du\;, 
\end{equation}
and returning to $z$-variable, the contribution of the angular momentum to the rotating black hole total energy yields
\begin{equation}
U_2(\omega)  = \lim_{\epsilon \to 0}  \left(  \int_{\epsilon}^{z_h} \rho^{(1)}_2(z, \omega) \,dz + \int_{ z_h}^{ \infty }\rho_2^{(2)}(z, \omega) \,dz  \right) \;,
\end{equation} 
whereby
\begin{eqnarray}\label{density2}
\rho^{(1)}_2(z,\omega) &=& - \frac{4L^3}{\kappa^2\gamma}   \frac{l^2\omega^2 \gamma^2\sqrt{c}}{z_h^4\sqrt{\pi}}  e^{-c z^2} \;, \quad \,\,\,\,\;\;\;\;\,\;\;\,\,\quad\quad\quad\quad \text{if} \quad \epsilon  \leq z \leq z_h\;, \nonumber\\
\rho^{(2)}_2(z,\omega) &=& \frac{4L^3}{\kappa^2\gamma}l^2 \omega^2 \gamma^2 \left(-  \frac{ \sqrt{c}}{z_h^4\sqrt{\pi}} +\frac{4 }{z^5}  +  \frac{2c }{z^3} \right) e^{-cz^2} \;, \quad \text{if} \quad z \geq z_h\;.
\end{eqnarray} 

Finally, one concludes that the total energy density of the cylindrical rotating BH in the grand canonical ensemble, obtained by adding the partial densities \eqref{density1} and \eqref{density2} is given by the expression

\begin{equation}\label{totalBHdensity}
\rho^{BH}(z,\omega) =
\begin{cases}
- \frac{4L^3}{\kappa^2\gamma} \frac{\sqrt{c}}{ z_h^4 \sqrt{\pi}}\left(\frac{3}{4}+l^2\omega^2\gamma^2 \right)e^{-cz^2} \;, & \quad\quad\quad\quad  \,\,\,\;\;\;\;\;\;\;\, \text{if} \quad \epsilon  \leq z \leq z_h\;,  \\ \\
  \frac{4L^3}{\kappa^2\gamma}\Big[ -\frac{\sqrt{c}}{z_h^4\sqrt{\pi}}\left(\frac{3}{4}+l^2\omega^2\gamma^2\right)+ \frac{1 }{z^5}\left(3+4l^2 \omega^2\gamma^2\right) \\ 
+\  \frac{2c }{z^3}\left(1+ l^2 \omega^2 \gamma^2\right)\Big] e^{-cz^2}\;,&   \quad \quad\quad\quad\quad\quad\;\;  \text{if} \quad z > z_h\;.
\end{cases}
\end{equation}

\section{Configuration entropy }

The configuration entropy (CE) definition is motivated by information theory. In particular in the Shannon information entropy \cite{shannon} that for a discrete random variable with probabilities $p_n$ of assuming one  of  $n$ possible values  is defined as:
 \begin{equation}
 S = - \sum_n  \, p_n  \log p_n \; .
 \label{discretepositionentropy}
 \end{equation}
The Shannon entropy represents a measure of the information content of the random variable. 

 A definition for the CE was proposed in references \cite{Gleiser:2011di,Gleiser:2012tu,Gleiser:2013mga} as a continuous version of Eq. \eqref{discretepositionentropy}. For a one-dimensional system it reads:    
  \begin{equation}
 S_C[f]  = - \int dk  \, f({k }) \ln f({k }) \,,
 \label{positionentropy}
 \end{equation}
 being $ f (k) $ the so-called modal fraction, usually defined, for a localized physical system,   in terms of the energy density in momentum space, $ \rho (k) $, namely   
 \begin{equation}
  f(k) = \frac{\vert  \rho (k)\vert^2}{\vert  \rho(k)\vert^2_{max}  } \,  \label{modalfraction}
 \end{equation}
 where $  \vert  \rho(k)\vert^2_{max} $ is the maximum value assumed by  $ \vert  \rho (k)\vert^2 $. Instead of the maximum value of the energy density, one could eventually use  $\int \vert  \rho (k)\vert^2  dk $ in the denominator of Eq. \eqref{modalfraction}. In this alternative definition, the modal fraction appears as a normalized function, which would be more similar to the Shannon entropy, but in the continuous case it could lead to negative values for the CE.

 It is now known through many examples, as those articles cited in the introduction, that the CE  works  as a measure of  the stability of physical systems. In a few words, the current interpretation states that the CE increases as the instability of the system increases.

 \subsection{ CE definition for the rotating QGP}
 
 The Fourier transform of the BH energy density is given by:
 \begin{eqnarray}\label{rho(k)}
 \widetilde{\rho}(k, \omega) = \frac{1}{2\pi} \lim_{\epsilon \to 0}   \int_\epsilon^{ \infty} dz\, \rho^{BH}(z, \omega) \,e^{i k z}    \;,
\end{eqnarray} 
with $\rho^{BH} (z, \omega)$    defined by Eq. \eqref{totalBHdensity}, where one notices that the total energy density has two different expressions, separated by the horizon position.

It is important to remark that the AdS black hole is a five-dimensional object. However, the energy density depends only on the coordinate $z$. There is a symmetry with respect to the other coordinates. In the holographic description, 
all the thermal properties of the plasma are encoded on the coordinate $z$. For example, the temperature is related to the location $ z = z_h$ of the BH horizon. So, the holographic description works as a one dimensional system and the Fourier transform is one dimensional with a single $k$ variable as the transformed of $z $. It is worth to stress the fact that the variable k has no relation to a 4D momentum of the plasma. It is just the Fourier transformed variable associated with the holographic coordinate $z $,  that does not exist in the plasma.

The modal fraction, necessary for building the CE, is defined in terms of the squared absolute value of $\widetilde{\rho}(k, \omega)$. Using \eqref{rho(k)}, it can be written as 
\begin{equation}
    \vert \widetilde{\rho}(k, \omega) \vert^2 = \left[ \frac{1}{2\pi} \lim_{\epsilon \to 0}  \int_\epsilon^{\infty} \rho^{BH} (z, \omega) \cos(kz)\,dz   \right]^2  
+ \left[ \frac{1}{2\pi} \lim_{\epsilon \to 0}   \int_\epsilon^{\infty} \rho^{BH} (z, \omega) \sin(kz)\,dz  \right]^2 \;.
\label{quadraticfourier}
\end{equation}
Eq. \eqref{quadraticfourier} above does not possess an analytical solution. So we apply numerical methods in order to determine the CE.

Using numerical integration, we plot in Figure 1 
$ \vert \widetilde{\rho}(k, \omega) \vert^2$ at $\bar{T} = 0.6$, where $\bar{T} = T/\sqrt{c}$. The IR parameter  $\sqrt{c}$  is  determined by hadronic phenomenology -- see, for instance, \cite{Herzog:2006ra}.
For other temperatures the pattern of the curve is similar.   

\begin{figure}[!htb]\label{densidadeBH_T6}
	\centering
	\includegraphics[scale=0.7]{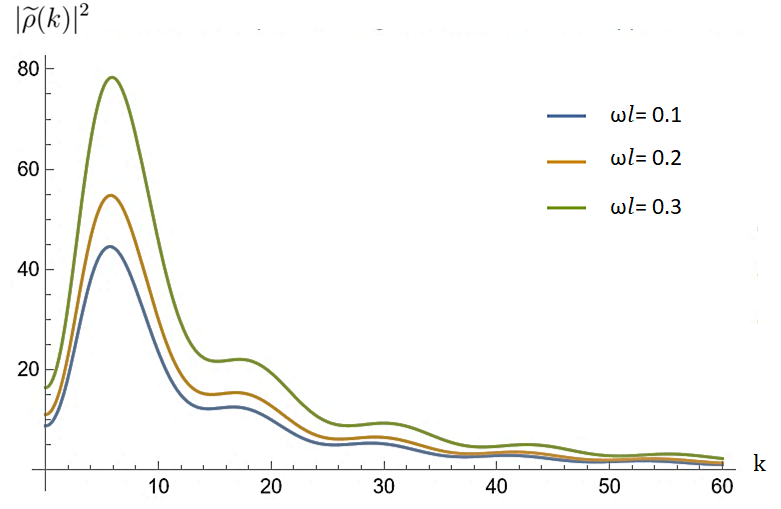}
	\caption{ Absolute value of the rotating BH energy density, $\vert \widetilde{\rho}(k) \vert^2$, versus momentum at $\bar{T}=0.6$ and different rotational velocities: $\omega l = 0.1$ (blue), $\omega l= 0.2$ (orange); and $\omega l = 0.3$ (green).}
\end{figure}

As one can see, $ \vert \widetilde{\rho}(k, \omega) \vert^2$ first reaches the global maximum $\vert \widetilde{\rho}(k) \vert^2_{max}$, then the curve begins to decrease as the momentum increases, oscillating smoothly and tending to zero as $k \rightarrow \infty$. Moreover, $\vert \widetilde{\rho}(k) \vert^2_{max}$ gets larger as the rotational speed increases, with a small increase in the value of $k$ where these maxima occur. These global maxima are well defined, and can be easily computed by numerical methods. After determining the maxima of the absolute value of the energy density in Fourier space \eqref{quadraticfourier}, we are able to evaluate the configuration entropy of the rotating plasma at different angular velocities and temperatures, by using the definition \eqref{positionentropy} together with \eqref{modalfraction}.

\subsection{ Results obtained for the CE }

Applying numerical methods, we computed the CE for different values of the dimensionless temperature $ \tilde T$ and of the rotational speed $ \omega l $. The values obtained are displayed in Table 1, in Appendix A. In order to show the behaviour of the CE we plot, in  Figure 2,  the case of fixed temperature $\bar{T}_6 = 0.6$ as a function of the rotational speed $\omega l $. One notices that the CE increases monotonically with $ \omega l $.  In particular, one also notices that when $\omega l $ approaches the speed of light: $\omega l \rightarrow 1$, the CE diverges. This behaviour is present for all the analyzed temperatures, as can be seen in  Figure 3, where we plot the CE for four different temperatures. In order to compare the variation of the CE with temperature and the variation with rotational speed, we plot in Fig. 4 the CE as a function of $ \tilde T$, at three different values of $\omega l$.  One notices that for a fixed  rotational speed, the CE increases with the temperature.

\begin{figure}[!htb]\label{CEvsT}
	\centering
	\includegraphics[scale=0.7]{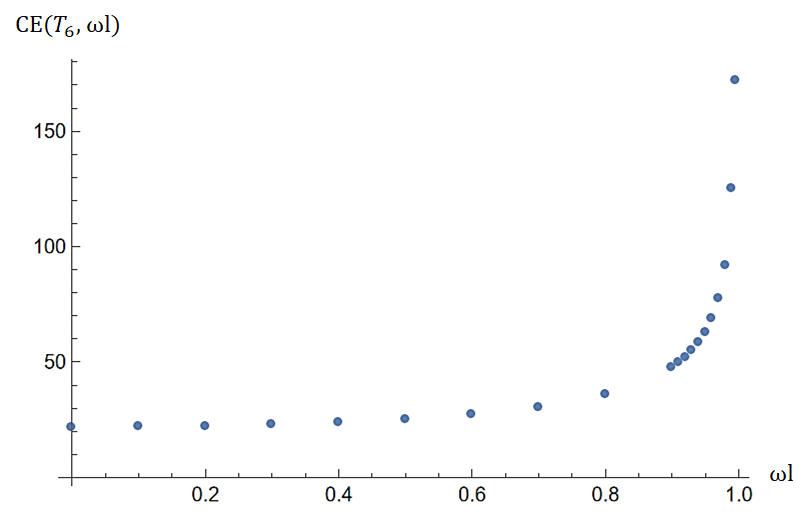}
	\caption{Configuration entropy of rotating QGP as a function of the rotational speed ($\omega l$) at $\bar{T}_6 = 0.6$.}
\end{figure}

\begin{figure}[!htb]\label{CEvsT2}
	\centering
	\includegraphics[scale=0.7]{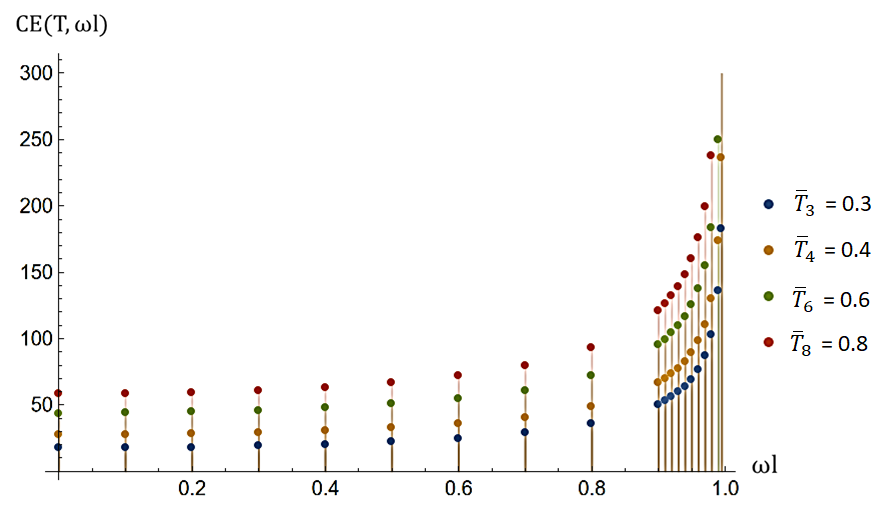}
	\caption{CE of rotating QGP as a function of $\omega l$, at different temperatures: $\bar{T}_3  = 0.3$ (blue), $\bar{T}_4 = 0.4$ (orange), $\bar{T}_6 = 0.6$ (green), and $\bar{T}_8 = 0.8$ (red).}
\end{figure}

\begin{figure}[!htb]\label{CEvsTfig}
	\centering
	\includegraphics[scale=0.65]{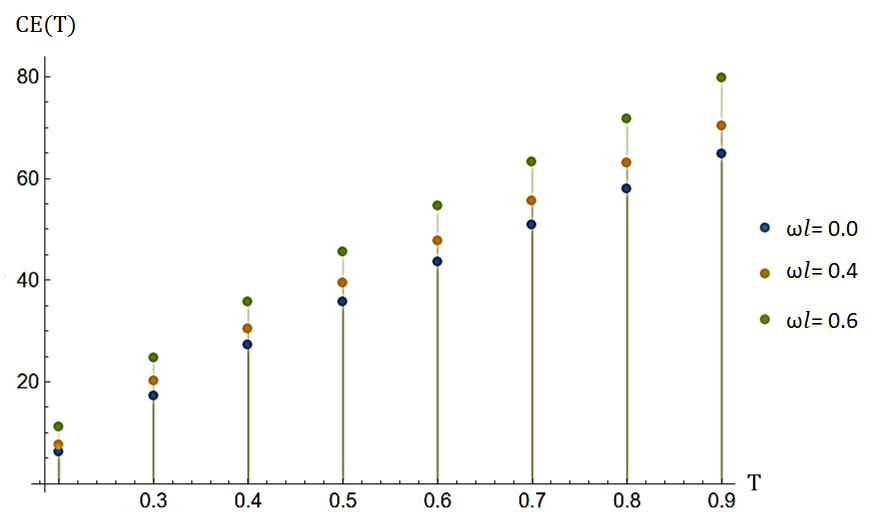}
	\caption{Configuration entropy of rotating QGP as a function of $\bar{T}$, at different rotational velocities: $\omega l  = 0$ (blue), $\omega l = 0.4$ (orange); and $\omega l = 0.6$ (green).}
\end{figure}

Increase in the CE is in general associated with increase in instability. However, as we have discussed in section {\textbf{2}}, rotation makes the plasma more stable against Hawking-Page transitions. So, what kind of instability could be associated with the increase in $\omega l $? 

In order to explore this point one can  
return to Equations \eqref{DeltaE}, \eqref{DeltaEBH} and \eqref{DeltaEAdS} for the action densities and perform the integral in the $z$ coordinate. The result is
\begin{equation}
	\bigtriangleup \mathcal{E}(\omega l , z_h ) \, = \,\frac{ L^3 \pi }{\kappa^2  z_h^3 } \bigg[  e^{-c z_h^2} ( - 1 + c z_h^2 ) + \frac{1}{2} + 
	c^2 z_h^4 \text{Ei}\left( -c z_h^2 \right) \bigg] \,,\label{FreeSW}
\end{equation}
that can be expressed in terms of the temperature, or its inverse, $\beta$,  using Eq. \eqref{HT}: 
\begin{equation}
    z_h = \frac{\sqrt{1-\omega^2 l^2}}{\pi T} = \frac{ \beta \sqrt{1-\omega^2 l^2}}{\pi }\,.
\end{equation}
It is very simple to show that the mass (energy) density of the black hole that emerges from Eqs. \eqref{totalE/DV} and \eqref{FreeSW} increases with the rotational speed, and diverges when $\omega l \to 1$. A result that is consistent with the fact that the energy of  objects, with non-vanishing rest mass, increase with their speed and goes to infinite when it goes to the speed of light. The black hole mass can be determined by the expression
\begin{equation}\label{MBH}
M = \Phi_G + sT + \omega J\;,     
\end{equation}
where $\Phi_G$ is the Gibbs free energy, $s$ is the BH entropy, and $J$ its angular momentum, given by
\begin{equation}\label{PhisJ}
    \Phi_G = \frac{1}{\beta} I_{total}\;, \quad s = - \frac{\partial \Phi_G}{\partial T}\;, \quad J = -\frac{\partial \Phi_G}{\partial \omega}\;; 
\end{equation}
wherein $I_{total} \equiv \bigtriangleup \mathcal{E}$ is the total energy density\footnote{In our case, it is implicit that $M$, $\Phi_G$, $s$ and $J$ were divided by the trivial volume of the bulk. }, see Eq. \eqref{FreeSW}. Thus, from Equations \eqref{MBH} and \eqref{PhisJ}, one obtains 
\begin{eqnarray}
 M &=&   \frac{L^3\gamma^5}{2\kappa^2}\left[ 3\pi^4T^4 \left( -1 + 2e^{-\frac{c(1-l^2\omega^2)}{\pi^2T^2}}\right)+2c\pi^2T^2(1-l^2\omega^2)e^{-\frac{c(1-l^2\omega^2)}{\pi^2T^2}}\right. \nonumber\\
 &+&\left. 2c^2(1-l^2\omega^2)^2\text{Ei}(-\frac{c(1-l^2\omega^2)}{\pi^2T^2})\right]\;, 
\end{eqnarray}
that increases monotonically with the rotational speed and diverges in the limit $\omega l \rightarrow 1$, as shown in Figure 5. The increase in the mass density of the black hole leads to an increase in the intensity of the Hawking radiation of the BH. In the limit of infinite mass density,  the loss of energy by radiation also diverges. 

A black hole with positive specific heat, like in our case, can stay in thermal equilibrium with a thermal radiation background. In this situation the loss of energy of the BH by radiation is compensated by the absorption of energy from the external radiation background.   However, in our holographic model there is no such external background but rather just the black hole. And the black hole is loosing energy by radiating, with an intensity that increases with the rotational speed. So, the rate of energy loss increases 
with $\omega l $, and this could be associated with an increase in instability.


A natural question to be made at this point:  is this holographic model appropriate for describing the QGP formed in heavy-ion collisions? Or, in other words, is it expected that for the actual plasma formed in collisions one could associate some increase in instability when the rotational speed increases? Instability of a system is in general associated with the fact that it can decay to some other state. This is the case of the QGP, since it is only in approximate thermal equilibrium during a finite time, after which the hadronic matter hadronizes and the plasma disappears. The QGP emits photons during its lifetime (see for example\cite{Wang:2001xh,Wang:2000pv,Boyanovsky:2003qm}), so it is radiating and losing energy.   The radiation intensity increases with the temperature and, since it is generated by the charges of the plasma, it is natural to expect that it also increases with the rotational speed, that corresponds to an increase in the acceleration of the charges. So, the intensity of energy loss of the QGP initial state increases with the rotational speed. The plasma of deconfined quarks and gluons is formed because of the very high energy density produced in the collisions. When it loses more energy by radiation because of the rotation, the energy density decreases faster and the time to hadronization tends to be smaller. This is what we interpret as an increase in instability.
 So, our results are consistent with the interpretation of the CE as an indicator of the stability of physical systems.

\begin{figure}[!htb]\label{MT6}
	\centering
	\includegraphics[scale=0.7]{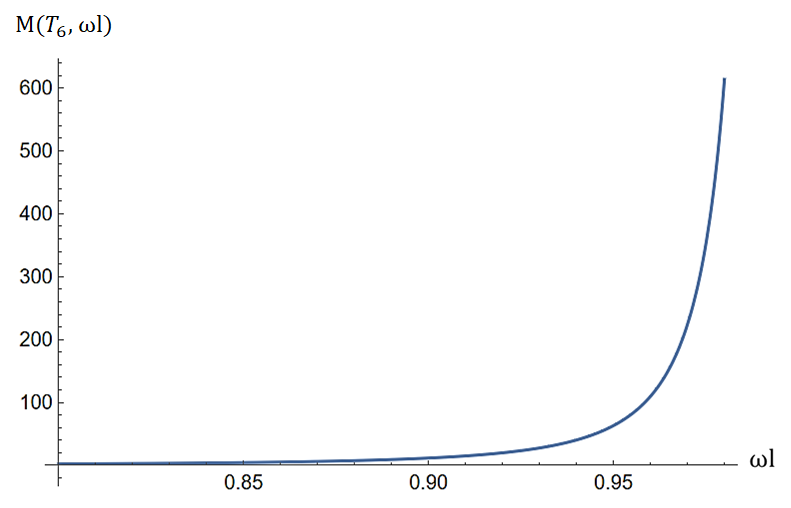}
	\caption{Mass of cylindrical rotating BH as a function of rotational speed ($\omega l$) at $T_6 = 0.6$, with $c=1$, and in units of $ L^3 /\kappa^2$.}
\end{figure}

\section{Conclusions}

We investigated the dependence of the configuration entropy on the rotational speed for a quark-gluon plasma with cylindrical symmetry using a holographic model. The plasma was represented  by the grand canonical ensemble (with null chemical potential). In this scenario, we obtained an expression for the energy density of the rotating AdS black hole, dual to the plasma,  that was applied to the calculation of the CE.  The dependence of  this quantity  on the rotational speed of the black hole was studied for different temperatures using numerical methods.

The current interpretation of the configuration entropy states that it works as a measure of the stability of physical systems. In short, the CE increases as the instability increases \cite{Gleiser:2011di,Gleiser:2012tu,Gleiser:2013mga}. The result found in section  4 -- the CE increases with the rotational speed $\omega l $ -- is consistent with this interpretation. Rotation of the plasma implies an increase in the mass density and angular momentum and, correspondingly, to an increase in the intensity of the emitted radiation. 

The effect of the variation of the temperature in the CE of an anti-de sitter black hole has been investigated in \cite{Braga:2016wzx}. It was found that, for a non-rotating plasma, the CE increases with the temperature indicating  the instability caused by the evaporation of the black hole via Hawking radiation, since the holographic model does not contain an external background that could provide energy to the BH. In the corresponding gauge theory side of the duality, the plasma  
 is loosing energy by emitting photons and/or some other processes. Such a description is in harmony with the results obtained in \cite{Wang:2001xh,Wang:2000pv,Boyanovsky:2003qm}, concerning the photon production from a thermalized quark-gluon plasma at RHIC energies.

\newpage

\section* {  Appendix} 
\appendix

\section{  CE$(T, \omega)$ }

\begin{table}[h]
\centering
\begin{tabular}[c]{|c|c|c|c|c|c|c|c|}
\hline 
 \textbf{X} & $\bar{T}_3 = 0.3$ & $\bar{T}_4 = 0.4$   & $\bar{T}_5 = 0.5$ & $\bar{T}_6 = 0.6$ & $\bar{T}_7 = 0.7$ & $\bar{T}_8 = 0.8$ & $\bar{T}_9 = 0.9$  \\
\hline
$\,\,\,\omega l = 0.0\,\,\,$ & $\,\,\,17.3177\,\,\,$ & $\,\,\,27.1699\,\,\, $ & $\,\,\,35.7645\,\,\,$ &$\,\,\,43.5432\,\,\,$ & $ \,\,\, 50.9320\,\,\, $ & $\,\,\,57.9978\,\,\,$ & $\,\,\,64.8159\,\,\,$ \\
\hline
$\,\,\,\omega l = 0.1\,\,\,$ & $\,\,\,17.4737\,\,\,$ & $\,\,\, 27.3514\,\,\, $ & $\,\,\,35.9698\,\,\,$ & $\,\,\,43.7748\,\,\,$ &$ \,\,\,51.1908\,\,\,  $ &$\,\,\,57.9978\,\,\,$ & $\,\,\,65.1263\,\,\,$ \\
\hline 
$\,\,\,\omega l = 0.2\,\,\,$ &$\,\,\,17.9544\,\,\,$ & $  27.9103$    & $\,\,\,36.6029\,\,\,$ &$\,\,\,44.4887\,\,\,$ &$51.9886 $ &$\,\,\,59.1616\,\,\,$ & $\,\,\,66.0833\,\,\,$\\
\hline
$\,\,\,\omega l = 0.3\,\,\,$ &$\,\,\,18.7997\,\,\,$ & $28.8953 $  & $\,\,\,37.7188\,\,\,$ &$\,\,\, 45.7479\,\,\,$ & $ 53.3956 $ & $\,\,\,60.7113\,\,\,$ & $\,\,\,67.7703\,\,\,$  \\
\hline
$\,\,\,\omega l = 0.4\,\,\,$ & $\,\,\,20.0886\,\,\,$ & $ 30.3998 $  & $\,\,\,39.4261\,\,\,$ &$\,\,\,47.6759\,\,\,$ & $ 55.5495 $ & $\,\,\,63.0780\,\,\,$ & $\,\,\,70.3519\,\,\,$  \\ 
\hline
$\,\,\,\omega l = 0.5\,\,\,$ &$\,\,\,21.9659\,\,\,$ &$ 32.5914 $    & $\,\,\,41.9220\,\,\,$ &$\,\,\,50.4976\,\,\,$ & $ 58.7014  $ & $\,\,\,66.5529\,\,\,$ & $\,\,\,74.1494\,\,\,$\\
\hline 
$\,\,\,\omega l = 0.6\,\,\,$ &$\,\,\,24.7137\,\,\,$ &$   35.7812 $   & $\,\,\,45.5766\,\,\,$ &$\,\,\,54.6347\,\,\,$ & $ 63.3233  $ & $\,\,\,71.6411\,\,\,$ &$\,\,\,79.6887\,\,\,$\\
\hline
$\,\,\,\omega l = 0.7\,\,\,$ &$\,\,\, 28.8948\,\,\,$ &$ 40.6155 $  &  $\,\,\,51.1591\,\,\,$ &$\,\,\,60.9631\,\,\,$ &$  70.3972 $ &$\,\,\,79.4415\,\,\,$& $\,\,\,88.1499\,\,\,$   \\
\hline
$\,\,\,\omega l = 0.8\,\,\,$ &$\,\,\,35.8003\,\,\,$ & $48.7337 $ & $\,\,\,60.6088\,\,\,$ &$\,\,\, 71.6941\,\,\,$ & $82.3997 $ &$\,\,\,92.6117\,\,\,$& $\,\,\,102.292\,\,\,$ \\ 
\hline
$\,\,\,\omega l = 0.9\,\,\,$ & $\,\,\,50.3687\,\,\,$ & $ 66.3365 $    & $ \,\,\,81.2459\,\,\,$ &$\,\,\,95.0799\,\,\,$ &$ 108.213$ &$\,\,\,120.801\,\,\,$ & $\,\,\,133.186\,\,\,$ \\
\hline 
$\,\,\,\omega l = 0.91\,\,\,$ & $\,\,\,52.9314\,\,\,$ & $ 69.4661 $    & $ \,\,\,84.9202\,\,\,$ &$\,\,\,99.1757\,\,\,$ &$ 112.752$ &$\,\,\,125.940\,\,\,$ & $\,\,\,138.910\,\,\,$ \\
\hline 
$\,\,\,\omega l = 0.92\,\,\,$ & $\,\,\,55.9181\,\,\,$ & $ 73.1201$ & $ \,\,\,89.2013\,\,\,$ &$\,\,\,103.917\,\,\,$ &$ 118.083$ &$\,\,\,131.964\,\,\,$ & $\,\,\,145.649\,\,\,$ \\
\hline 
$\,\,\,\omega l = 0.93\,\,\,$ & $\,\,\,59.4664\,\,\,$ & $ 77.4686 $    & $ \,\,\,94.2683\,\,\,$ &$\,\,\,109.542\,\,\,$ &$ 124.557$ & $\,\,\,139.232\,\,\,$ & $\,\,\,153.798\,\,\,$\\
\hline
$\,\,\,\omega l = 0.94\,\,\,$ & $\,\,\,63.7885\,\,\,$ & $ 82.7700 $    & $ \,\,\,100.376\,\,\,$ &$\,\,\,116.435\,\,\,$ &$ 132.489$ & $\,\,\,148.253\,\,\,$ & $\,\,\,163.931\,\,\,$ \\
\hline 
$\,\,\,\omega l = 0.95\,\,\,$ & $\,\,\,69.2319\,\,\,$ & $ 89.4363 $    & $ \,\,\,107.993\,\,\,$ &$\,\,\,125.228\,\,\,$ &$ 142.681$ & $\,\,\,159.877\,\,\,$ & $\,\,\,177.005\,\,\,$ \\
\hline 
$\,\,\,\omega l = 0.96\,\,\,$ & $\,\,\,76.4179\,\,\,$ & $ 98.1476 $    & $ \,\,\,118.090\,\,\,$ &$\,\,\,137.079\,\,\,$ &$ 156.494$ & $\,\,\,175.663\,\,\,$ & $\,\,\,194.774\,\,\,$ \\
\hline 
$\,\,\,\omega l = 0.97\,\,\,$ & $\,\,\,86.5942\,\,\,$ & $110.284 $    & $ \,\,\,132.700\,\,\,$ &$\,\,\,154.408\,\,\,$ &$ 176.772$ & $\,\,\,198.872\,\,\,$ & $\,\,\,220.913\,\,\,$\\
\hline 
$\,\,\,\omega l = 0.98\,\,\,$ & $\,\,\,102.651\,\,\,$ & $ 129.994 $    & $ \,\,\,157.074\,\,\,$ &$\,\,\,237.976\,\,\,$ &$ 210.908$ & $\,\,\,237.976\,\,\,$ & $\,\,\,264.957\,\,\,$ \\
\hline 
$\,\,\,\omega l = 0.99\,\,\,$ & $\,\,\,136.231\,\,\,$ & $ 173.956 $    & $ \,\,\,212.166\,\,\,$ &$\,\,\,249.391\,\,\,$ &$ 288.362$ & $\,\,\,326.702\,\,\,$ & $\,\,\,364.847\,\,\,$ \\
\hline 
$\,\,\,\omega l = 0.995\,\,\,$ & $\,\,\,182.902\,\,\,$ & $ 236.174 $    & $ \,\,\,290.410\,\,\,$ &$\,\,\,342.816\,\,\,$ &$ 398.298$ & $\,\,\,452.576\,\,\,$ & $\,\,\,506.449\,\,\,$ \\
\hline 
\end{tabular}   
\caption{Configuration entropy of rotating QGP at different temperatures ($\bar{T} = T/\sqrt{c}$) and rotational velocities ($\omega l$).}
\label{table1}
\end{table} 

\noindent \textbf{Acknowledgments}: The authors are supported by FAPERJ --- Fundação Carlos Chagas Filho de Amparo à Pesquisa do Estado do Rio de Janeiro, CNPq - Conselho Nacional de Desenvolvimento Cient\'ifico e Tecnol\'ogico. This work received also support from  Coordena\c c\~ao de Aperfei\c coamento de Pessoal de N\'ivel Superior - Brasil (CAPES) - Finance Code 001. We also wish to acknowledge the financial support provided by ANID (Chile) under Grants No.3220304 (L. F. F.).

\bibliographystyle{utphys2}
\bibliography{library}
\end{document}